# A Genetic Algorithm based Approach for Test Data Generationin Basis Path Testing

Yeresime Suresh
Department of Computer Science and Engineering,
National Institute of Technology,
Rourkela 769008, India
suresh.yeresime@gmail.com

Santanu Ku Rath
Department of Computer Science and Engineering,
National Institute of Technology,
Rourkela 769008, India
skrath@nitrkl.ac.in

*Abstract*—Software testing is a process to identify the quality and reliability of software, which can be achieved through the help of proper test data. However, doing this manually is a difficult task due to the presence of huge number of predicate nodes in the module. So, thisleads towards a problem of NP-complete. Therefore, someintelligence-based search algorithms have to be used to generate test data.In this paper, a soft computing based approach for generating test data automatically using genetic algorithmbased upon the set of basis paths is proposed.This paper combines the characteristics of genetic algorithm with test data, making use of the merits of respective global and local optimization capability to improve the generation capacity of test data.This automated process of generating test data optimally helps in reducing the test effort and time of a tester.Finally, the proposed approach is applied tothe ATM withdrawal task. Experimental results show that GA was able to generate suitable test data based on a fitness value and avoid redundant data by optimization.

## I. INTRODUCTION

The process of automatic generation of test data plays a major role in software testing. Software testing is generally divided into white box testing and black box testing. White box testing is also known as structural testing and basis path testing is one among them in structural testing.The emphasis is on finding specific input data. Therefore, automatic generation of test data is one of the key research topics in software testing. Today, researchers as well as practitioners use more common methods such as notion to perform, random method and heuristic approaches for test data generation[1]. These methods have some pitfalls in generating test data for larger and complicated programs. So other intelligence techniques have been used very much.

Test data generation in program testing, is the process of identifying a set of test data, which satisfies the given testing criterion[2]. A test data generator is a tool which helps a tester in generation of test data for a given program. Most of the existing test data generators have been classified into three types viz., path wise test data generators[3,4,5,6,7], Data specification generators[8,9,10,11] and random test data generators [12], however practically these techniques require complex algebraic computations.

In this paper, we discuss Genetic algorithm (GA), anon-traditional approach for generating test data based on a set of basis paths. All the paths selected in a module need to be executed and thus generating a large set of test data for these pathsis quite a difficult task. As a result, certain degree of automated process should be carried out to minimize testing resources. GA helps in achieving this goal by optimizing test data required to cover the paths in a control flow graph.

The rest of the paper is organized as follows: Section 2 presents the need for test data automation; Section 3 gives an Overview of Basis path. Section 4 represents the fundamentals of GA for test data generation. Section 5 presents the proposed approach for test data generation using GA. Section 6 represent the results, and section 7 concludes the paper.

## II. NEED FOR TEST DATA AUTOMATION

Testing is defined as the process of executing a program with the intent of finding errors[13].Software testing can also be defined as a process, or a series of processes, designed to make sure the code does what it was designed to do and that it does not do anything unintended[13]. Software should be predictable and consistent, offering no surprises to users. The main objective of testing is to prove that the software product as a minimum meets a set of pre-established acceptance criteria under a prescribed set of environmental circumstances. There are two components to this objective. The first component is to prove that the requirements specification from which the software was designed is correct. The second component is to prove that the design and coding correctly respond to the requirements[14]. Automatic generation of test data helps in reduction of execution time and discovering errors. Automating the process of test data generation reduces the cost in developing test cases.

## III. OVERVIEW OF BASIS PATHS

Basis path testing is one of the famous structural testing criteria[15]. It is a methodology which searches the program domain for suitable test data, such that after executing the program with the test data, a predefined path is reached. Based on the cyclomatic complexity analysis, each path is being tested for efficient functionality. Practically it is possible to apply path testing for a specific subset of paths in the control flow graph. This mechanism aims to compute the logical complexity of a procedural design and defines a set of execution paths. Test data are generated in such a way that they will execute every statement at least once.

Cyclomatic Complexity[15]is used to evaluate the complexity of an algorithm in a method. It is an indication of thenumber of test cases that are required to test the method completely, which helps a tester to estimate the number of test cases required to achieve maximum code coverage. Cyclomatic complexity is a measure of the logical complexity of a module, and the minimum effort necessary to qualify a module. Cyclomatic complexity is the number of linearly independent paths, and consequently the minimum number of paths that one shouldtest. Based on the cyclomatic complexity measure, structured testing uses the control flow graph of software to establish path coverage criteria. The resultant test sets provide more thorough testing than statement and branch coverage. Cyclomatic complexity is given by the equation $V(G) = e - n + 2$, where 'e' and 'n' are the number of edges and number of nodes in a control flow graph(CFG) respectively. The value of V (G) is an indication of all the possible paths of execution in the program and these set of paths are referred to as basis paths. A basis path is defined as a sequence of instructions or statements that start at an entry point and ends at another, or possibly the same, or exit. The value of V (G) implies a lower bound on the number of test cases required to test the method completely.Table-1 shows the list of literature survey and the criteria used by various authors in basis path testing.

**Table-1**: Literature survey for test data generation in basis paths using GA.

| Author | Criteria for Testing |
|---|---|
| Duran J.W, Ntafos S.C | Random Testing: Segment, Branch Coverage[16]. |
| DeMillo R, Offutt A.J | Generating test case values using adequacy based testing criteria[17]. |
| L. Clarke | Path coverage testing criteria' to generate the test data. He selected target paths, executed them, and then generated test data such that the identified constraints are satisfied[5]. |
| Bogdan Korel | Path testing : dynamic path testing technique that generates test data by executing the program with different possible test data values[2]. |
| Mansour N, Salame M | path coverage testing criteria for generating test data using hamming distance as a fitness function[18]. |
| Srivastava P.R, Kim T | Focused on path coverage testing criteria and proposed a technique for generating test cases using GA[20],emphasizing on the critical paths during testing. |
| Michael et al | Branch coverage: automated test data generation.[21]. |
| Wegener et al. | Structural test coverage criteria to generate test data by using evolutionary approach[22]. |
| Lin J.C, Yeh P.L | Path testing:automated test data generation [19]. |
| Xanthakis | Search based software test data generation using heuristic search procedures[23]. |
| Rauf A, Anwar S | GUI based test criteria: generate test data using GA's. Sequences of events represent the candidate test case values.Number of paths followed out of the total number of paths was used as a fitness function[24]. |
| McMinn P | Search based software test data generation[25]. |
| Ahmed M.A | Path coverage criteria to generate test data using GA[26]. |
| Shen et al. | Proposed GATS algorithm, which is a hybrid scheme of GA and Tabu search, to generate test data. Focus was on Function coverage testing criteria[27]. |
| Harman M | Search based software engineering for automated test data generation[28]. |
| Malhotra et al | Test data generation using machine learning techniques for the object oriented software[29]. |

IV. GENETIC ALGORITHM

GAis an optimization and machine learning algorithm based loosely on the processes of biological evolution. John Holland created the GA field [30] and it is the first major GA publication. GA provides a general-purpose search methodology, which uses the principles of natural evolution [31].

Genetic algorithm as an effective global smart search method, reveals its own strength and efficiency to solve the large space, optimized for high complicated problems, and thus provides a new method to solve the problems of generating test data [1].GA solves optimization problems by manipulating initial population (individual chromosomes sampled randomly). Each chromosome is evaluated based on a fitness function which is related to its success in solving a given problem. Given an initial population of chromosomes, GAproceeds by choosing chromosomes to serve as parents and then replacing members of the current population with new chromosomes that are (possibly modified) copies of the parents. The process of selectionand population replacement goes on until a stopping criterion (achieving effective test data) has been met [32].

Thus,GA has been successfully used to automate the generation of test data. GA begins with a set of initial population which is randomly sampled for a particular problem domain. Then GA is applied, by performing a set of operations iteratively to get a new and fitter generation. Generating test data automatically reduces the time and effort of the tester.

The two common operations that are performed to produce efficient solution for a target problem after selection operation are Crossover and Mutation.

   a. Crossover

This operation is used to produce the descendants that make up the next generation. This operation involves the following cross breeding procedures[19].

i. Randomly select two individuals as a couple from the parent generation.
ii. Randomly select a position of the genes, corresponding to this couple, as the cross over point. Thus, each gene is divided into two parts.
iii. Exchange the first parts of both genes corresponding to the couple.
iv. Add the two resulted individuals to the next generation.

b. Mutation

The mutation operation picks a gene at random and changes its state according to the mutation probability. Mutation maintains diversity in a generation to prevent premature convergence to a local optimal solution.

Mutation operation is carried out after Crossover. Mutation is an operation in which the chromosomal bit representation of zero's are flipped into one's and vice versa based on the mutation probability($p_m$). GA guarantee high probability of improving the quality of the individuals over several generations according to the Schema Theorem[31]. Mutation generally prevents GA from falling into local extremes. Mutation shouldn't occur frequently, because GA will change into random search.

*A. Fitness Function for Basis Paths*

A fitness function for test data generation for an ATM withdrawal task is developed based on Bogdan Korel's branch distance function [2]. Consider a path 'P' in the program execution. The goal of the test data generation problem is to find a program input 'x' on which P will be traversed. Without loss of generality, Korel assumed that the branch predicates are simple relational expressions (inequalities and equalities). That is, all branch predicates are of the form: E1 opE2, where E1 and E2 are the arithmetic expressions and op is one of $\{<, \leq, >, \geq, =, \neq\}$ the operator. In addition, he assumed that predicates do not contain AND's or OR's or other boolean operators. Each branch predicate E1 opE2 can be transformed to the equivalent predicate of the form F rel 0, where F and rel are given in Table-2.

**Table-2**: Equivalent predicate of branch function.

| Branch Predicate | Branch Function F | rel |
|---|---|---|
| $E_1 > E_2$ | $E_2 - E_1$ | $<$ |
| $E_1 \geq E_2$ | $E_2 - E_1$ | $\leq$ |
| $E_1 < E_2$ | $E_1 - E_2$ | $<$ |
| $E_1 \leq E_2$ | $E_1 - E_2$ | $\leq$ |
| $E_1 = E_2$ | $abs(E_1 - E_2)$ | $=$ |
| $E_1 \neq E_2$ | $abs(E_1 - E_2)$ | $\leq$ |

F is a real valued function, referred to as branch function, which is 1) positive (or zero if rel is <) when a branch predicate is false or 2) negative (or zero if rel is = or ≤ ) when the branch predicate is true. It is obvious that F is actually a function program input. But this process requires a very large and complex algebraic manipulation. For this reason an alternative approach was used in which the branch function was evaluated, as basis path testing includes both statement testing and branch testing. For example to test "if a > b then…" has a branch function F, whose value can be computed for a given input by executing the program and evaluating 'a-b' expression.

This concept was used in our approach to test the ATM withdrawal task. We generated test data for a feasible basis path in theCFG. From CFG, we can compute the number of paths required to be tested.We have generated test data for a single feasible with respect to an ATM withdrawal task[33].

V. PROPOSED SYSTEM

The concept of GA has been applied to the problem of automated test data generation process. Here the test data is referred to as population in GA. In initial population, each individual bit string (chromosome) is a test data. This set of chromosomes is used to generate test data for feasible basis paths.

The system for generating automated test data for feasible basis paths using GA has been coded in MATLAB. It randomly generates the initial population, evaluates the individual chromosome based on the fitness function value and applies the GA operations such as selection, crossover and mutation to produce next generation. This iterative process stops when the GA finds optimal test data.

*A. Fitness function design for our approach*

We have taken up a case study, describing a customer's activity of withdrawing money from an ATM[33]. Each customer in the bank system has an account and an ATM debit card. The scenario considered here for design of fitness function is that the customer tries to withdraw certain amount from the ATM machine (this withdrawal amount is the initial test data generated randomly, with an assumption that customer entering the withdrawal amount is random).Figure-1 shows the sequence of operations performed in ATM withdrawal task by the customer.

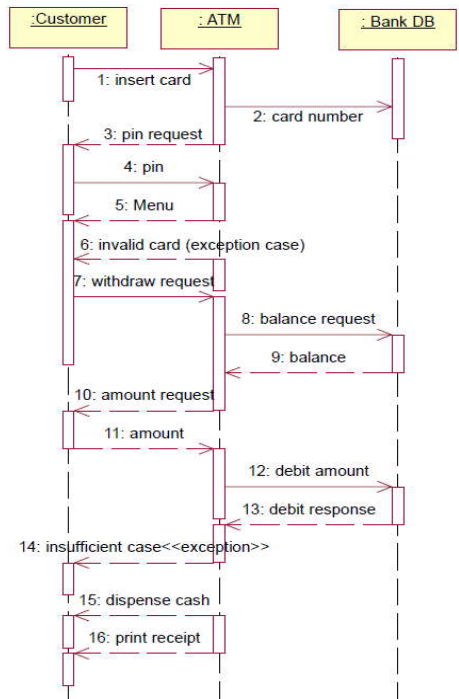

**Figure-1**: Sequence diagram for ATM Withdrawal.

The ATM system then sends the amount, and the account number to the bank system. The bank system retrieves the current balance of the corresponding account and compares it with the entered amount. If the balance amount is found to be greater than the entered amount then the amount can be withdrawn andthe bank system returns true, after which the customer can withdraw the money otherwise it checks for credit limit if the entered amount is less than the total amount (current balance) then return false. Depending on the return value, the ATM machine dispenses the cash and prints the receipt or displays the failure message.

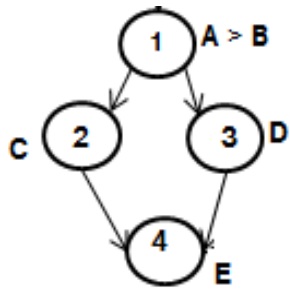

**Figure-2**: CFG for a sample code block.

The fitness function for the ATM withdrawal scenario was based on the traversal of predicate nodes. For instance, in Figure-2 when node-1 is visited the condition of the predicate node may be either 'A > B' or B > A or even A = B conditions may occur. So now taking equality condition into consideration, A = B implies A – B = 0; as GA for test data generation is minimization the fitness function 'f' is given as 1 / (A - B). But this functional value 'f' will evaluate to infinity when A – B=0, so to avoid this condition a small delta value is added to the fitness function. Hence the fitness function in general is given as: $f = 1 / ((abs (A - B) + 0.5)^2)$.

### B. Applying Genetic Algorithm for Path Testing

The principle of GAs has been applied to generate test data automatically. The developed system generates optimal test data automatically on the basis of basis paths in the control flow graph. The first generation is generated randomly and then by performing the basic GA steps, fitness of individuals gets improved. The system first generates the individual test data randomly, and then calculates fitness for each individual chromosome (test data) and on the basis of their fitness values it performs mutation and crossover. This process continues until all individuals reach to the maximum fitness. The system performs all operations from initial population to last generation automatically; it does not require the user interference. Generating test data automatically reduces the time and effort of the tester.

#### 1) Deriving test data based on Control flow graph

1. Using the source code of the program, draw the corresponding control flow graph (manually or automated).
2. Determine the cyclomatic complexity of the flow graph.
3. Determine the basis set of linearly independent paths.
4. Prepare test data that will force the execution of each path in the basis set.

This set of data generated randomly is the initial population Input) of the GA process to start.

The following lines of code indicate the ATM withdrawal task.

1. net_amt = 25000;
2. bal(1,i) = net_amt - wd_amt(1,i);
3. if wd_amt(1,i) < net_amt
4. if bal(1,i) < min_bal
5. fail_bal(1,k) = bal(1,i);
    else
6. suc_bal(1,p) = bal(1,i);
7. test_data(1,p) = wd_amt(1,i);

Control flow graph construction for ATM withdrawal task for source code shown above (target path) is shown in Figure-3.

8. Gen = Gen +1
9. go to Step 3
10. end
11. Select the chromosome having the best fitness value as the desired result (test data for target path).

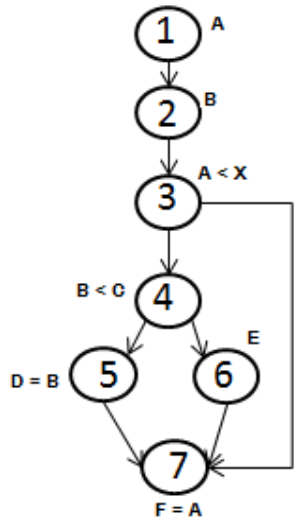

**Figure-3**: Control flow graph for ATM withdrawal.

**Table-3**: Alphabetical representation of Predicate Nodes in CFG for Fig. 3

*2) Genetic algorithm for test data generation*

The following steps show the algorithmic approach followed to generate test data for the basis path derived from CFG using GA. Figure-4 shows the schematic representation of test data generation using GA.
Algorithm:
Input: Randomly generated numbers (initial population act as test data) based on the target path to be covered.
Output: Test data for the target path.
1. Gen = 0
2. While Gen < 500
3. do
4. Evaluate the fitness value of each chromosome based on the objective function.
5. Use Elitism as selection operator, to select the individuals to enter into the mating pool.
6. Perform two-point cross over on the individuals in the mating pool, to generate the new population.
7. Perform bitwise Mutation on chromosomes of the new population

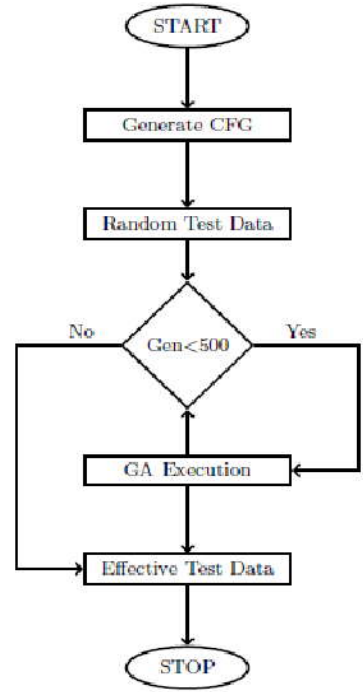

**Figure-4**: Basic flow of test data generation along with GA process.

| Predicate Nodes | Alphabetical Notation |
|---|---|
| wd_amt | A |
| net_amt | X |
| bal | B |
| min_bal | C |
| Fail | D |
| suc_bal | E |
| test_data | F |

*3) Experimental Settings*

The following sets of parameters were considered for test data generation using GA.
a. Fitness function :
   f = 1 / ((abs (suc_bal (i) - min_bal) + 0.05) ^2)
b. Coding : Binary String
c. Length of the string in the chromosome : 15bits
d. Population Size (N) :100
e. Selection method :Elitism
f. Two-point crossover and $p_c$ = 0.5
g. Mutation probability( $p_m$) = 0.05
h. Stopping Criteria = number of generation (500)

First set of test data was generated randomly. The test datathat we derived based on the set of basis paths, depends on the programs structure with an aim to traverse every executable statement in the program. The fitness function used was derived on the basis of branch distance[2]. The input variables were represented in binary form. The main objective of using GAs lies in their ability to handle input data which may be complex in nature. Thus, the problem of test data generation is treated entirely as an optimization problem. One of the merits of using GAs is that through the search and optimization process, test data sets are improved in a manner that they are close to the input domain.

VI. RESULTS

The approach followed for test data generation for path testing using GA, the following four basic steps were processed viz., Control Flow Graph Construction, Target Path Selection, Test Data generation and Execution, Test Result Evaluation.

Table-4 shows the fitness value range of test data and the classification of individual chromosome into their respective classes based on fitness value in terms of percentage.

**Table-4**: % Class of Test Data having maximum fitness value

| Fitness Value Range | % of Test Data |
|---|---|
| 0 ≤ f(x) < 0.3 | 61 |
| 0.3 ≤ f(x) < 0.7 | 01 |
| 0.7 ≤ f(x) < 1.0 | 38 |

Table-4 gives us a clear picture that around 38% of test data have higher fitness value 'f(x)' and lie in the range between 1.0 and 0.7. Figure-5 gives the graphical representation of test data in terms of percentage.

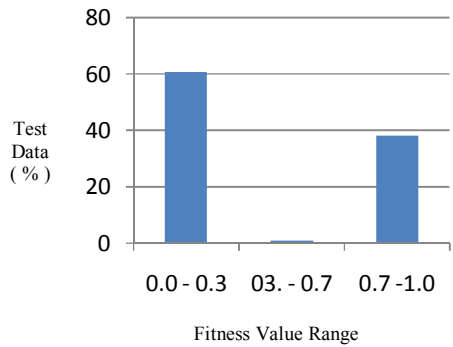

**Figure-5**: Graphical Representation of fitness value for Table-4.

VII. CONCLUSIONS

In software development life cycle, software testing is one of the critical phases. So generation of test data automaticallyis a key step which has a great influence on code coverage in software testing.

In this paper, a GA based on theory of natural selection was used to generate test data automatically for feasible basis paths. After the generation of initial test data randomly, GA was iterated for 500 generations as in practicality computation time should be finite. This paper makes use of a fitness function based on the condition of the predicate node.

The results in thispaper are an indication that GA is more effective and efficient in generating automated test data rather than random testing.

The future perspective of the work would be to enhance automated test data generation for large and complex programs, as of now the existing methods generate test data for smaller and simple programs.

Another prospective area of future study would be to generate test data using a fitness function for multiple paths in the control flow graph. The test data generated using GA can be used in code coverage analysis by comparing with other artificial intelligence techniques such as Particle swarm optimization, Simulated annealing, Clonal selection algorithm etc.

REFERENCES

[1] Wang Xibo and Su Na, "Automatic test data generation for path testing using genetic algorithms," in *Proc. 3rd International Conference on Measuring Technology and Mechantronics Automation ( ICMTMA )*, 2011, pp. 596-599.

[2] Bogdan Korel, "Automated Software Test Data Generation," *IEEE Transactions on Software Engineering*, vol. 16, no. 8, pp. 870-879, August 1990.

[3] Janis Bicevskis, Juris Borzovs, Uldis Straujums, Andris Zarins, and Edward F. Miller Jr., "SMOTL- A system to construct samples for data processing program debugging," *IEEE Transactions on Software Engineering*, vol. SE-5, no. 1, pp. 60-66, January 1979.

[4] R. Boyer, B. Elspas, and K. Levitt, "SELECT-A formal system for testing and debugging programs by symbolic execution," *SIGPLAN Notices*, vol. 10, no. 6, pp. 234-245, June 1975.

[5] L. Clarke, "A system to generate test data and symbolically execute programs," *IEEE Transactions on Software Engineering*, vol. SE-2, no. 3, pp. 215-222, September 1976.

[6] W. Howden, "Symbolic testing and the DISSECT symbolic evaluation system," *IEEE Transactions on Software Engineering*, vol. SE-4, no. 4, pp. 266-278, July 1977.

[7] C. Ramamoorthy, S. Ho, and W. Chen, "On the automated generation of program test data," *IEEE Transactions on Software Engineering*, vol. SE-2, no. 4, pp. 293-300, December 1976.

[8] J. Bauer and A. Finger, "Test plan generation using formal grammars," in *Proc. 4th International Conference on Software Engineering*, 1970, pp. 425-432.

[9] W. Jessop, I. Kanem, S. Roy, and J. Scanlon, "ATLAS - An automated software testing system," in *Proc. 2nd International Conference on Software Engineering*, 1976.

[10] N. Lyons, "An automatic data generation system for data base simulation and testing," *ACM SIGMIS Data Base*, vol. 8, no. 4, pp. 10-13, 1977.

[11] E. Miller Jr and R. Melton, "Automated generation of testcase datasets," *SIGPLAN Notices*, vol. 10, no. 6, pp. 51-58, June 1975.

[12] D. Bird and C. Munoz, "Automatic generation of random self-checking test case," *IBM System Journal*, vol. 22, no. 3, pp. 229-245, 1983.

[13] Glenford J. Myers, *The art of software testing*, 2nd ed.: Wiley, 2004.

[14] S. Kuppuraj and S. Priya, "Search Based Optimization for Test Data Generation Using Genetic Algorithms," in *Proc of the 2nd International Conference on Computer Applications*, 2012, pp. 201-205.

[15] Thomas J. McCabe, "A complexity measure," *IEEE Transactions on Software Engineering*, vol. 2, no. 4, pp. 308-320, December 1976.

[16] Duran J. W and Ntafos S. C, "An evaluation of random testing," *IEEE Transactions on Software Engineering*, vol. 10, no. 4, pp. 438-443, 1984.

[17] DeMill R and Jeff Offutt, "Constraint-based automatic test data generation," *IEEE Transactions on Software Engineering*, vol. 17, no. 9, pp. 900-910, September 1991.

[18] Mansour N and Salame M, "Data Generation for Path Testing," *Software Quality Journal*, vol. 12, pp. 121-136, 2004.

[19] Lin J. C and Yeh P. L, "Automatic test data generation for path testing using GA's ," *Information Sciences*, vol. 131, pp. 47-64, 2001.

[20] Srivastava P. R and Kim T, "Application of Genetic Algorithm in Software Testing," *International Journal of Software Engineering and Its Applications*, vol. 3, no. 4, pp. 87-96, 2009.

[21] Christoph C Michael, Gary McGraw, and Michael A. Schatz, "Generating software test data by evolution," *IEEE Transactions on Software Engineering*, vol. 27, no. 12, pp. 1085-1110, December 2001.

[22] Wegener J, Baresel A, and Sthamer H, "Evolutionary Test Environment for Automatic Structural Testing," *Information and Software Technology*, vol. 43, pp. 841-854, 2001.

[23] S, Xanthakis; C, Ellis; C, Skourlas; A, Le Gall; S, Katsikas; K, Karapoulios, "Application of genetic algorithm in software testing," in *Proceedings of 5th International Conference on Software Engineering and its Applications*, Toulouse, France, 1992, pp. 625-636.

[24] Rauf A and Anwar S, "Automated GUI Test Coverage Analysis using GA," in *Seventh International Conference on Information Technology*, 2010, pp. 1057-1062.

[25] McMinn P, "Search-based software test data generation: A survey," *Software Testing, Verification and Reliability*, vol. 14, no. 2, pp. 105-156, 2004.

[26] Moataz A. Ahmed and Irman Hermadi, "GA based multiple paths test data generator," *Computers & Operations Research*, vol. 35, pp. 3107-3124, February 2007.

[27] Xiajiong Shen, Qian Wang, Peipei Wang, and Bo Zhou, "Automatic Generation of Test Case based on GATS Algorithm," in *IEEE International Conference on Granular Computing, GRC'09*, 2009, pp. 496-500.

[28] Harman Mark, "Automated Test Data Generation using Search Based Software Engineering," in *Second International Workshop on Automation of Software Test (AST'07)*, 2007, pp. 1-2.

[29] Malhotra R and Garg M, "On the Applicability of Machine Learning Techniques for Object Oriented Software Fault Prediction," *Software Engineering : An International Journal (SEIJ)*, vol. 1, no. 1, September 2011.

[30] J. H. Holland, *Adaptation in Nature and Artificial Systems*.: Addison-Wesley, Reading, MA, 1975.

[31] D. E. Goldberg, *Genetic Algorithms in Search, Optimization and Machine Learning*.: Addision-Wesley, Reading, MA, 1989.

[32] James D. Kelly Jr and Lawrence Davis, "A Hybrid Genetic Algorithm for Classification," *International Joint Conference on Artificial Intelligence*, pp. 645-650, 1991.

[33] Michael R Blaha and James R Rumbaugh, *Object-oriented modeling and design with UML*, 2nd ed.: Pearson, 2005.